\documentstyle[12pt]{article}
\def\be{\begin{equation}}
\def\ee{\end{equation}}

\title{{\hfill\small\tt Lin.\,Alg.\,Appl.\,25,\,91-94\,(1979)}\\
{\hfill}\\
Some Diophantine relations involving circular functions of
rational angles}

\author{F. Calogero,\\
{\small\em Istituto di Fisica, Universit\`a di Roma}\\
{\small\em 00185 Roma, Italia}\smallskip\\
A.M. Perelomov \\
{\small\em Institute of Theoretical and Experimental Physics}\\
{\small\em B. Cheremushkinskaya, 25}\\
{\small\em 117259 Moscow, USSR}}

\date{}

\begin{document}
\maketitle

\begin{abstract}\noindent
The eigenvalues of the 3 off-diagonal matrices of rank $n$ with
elements $1+i\,\mbox{cot}\,[(j-k)\pi /n]$,
$\mbox{sin}^{-2}[(j-k)\pi /n]$ and $\mbox{sin}^{-4}[(j-k)\pi
/n]$, ($j=1,2,\ldots ,n$, $k=1,2,\ldots ,n$, $j\neq k$) are
computed. The sums over $k$ from 1 to $n-1$ of
$\mbox{cot}\, (k\pi/n)\,\mbox{sin}\,(2sk\pi /n)$
and $\mbox{sin}^{-p}(k\pi/n) \mbox{cos} (2sk\pi/n)$
 are moreover computed for $s$ integer and $p$=2 and 4.
The results are given by simple formulae in terms of integers.
\end{abstract}

\noindent In this note we report some remarkable relations
involving circular functions of rational angles.

\noindent
{\bf Theorem 1.} {\em The off-diagonal hermitian matrix of rank} $n$
{\em whose elements are defined by the formula}
\begin{equation}
A_{jk}=(1-\delta _{jk})\,\left\{ 1+i\,\mbox{cot}\,\frac{(j-k)\pi }{n}\right\}
\end{equation}
{\em has the integer eigenvalues}
\begin{equation}
a_s=2\,s- n -1,\qquad s=1,2,\ldots ,n \end{equation}
{\em and the corresponding eigenvectors} $v^{(s)}$ {\em have components}
\begin{equation}
v_j^{(s)}=\mbox{exp}\left( -\,\frac{2\pi i sj}{n}\right) ,\qquad
j=1,2,\ldots ,n. \end{equation}

\medskip
\noindent
{\bf Theorem 2.} {\em The } 2 {\em off-diagonal hermitian matrices whose
elements are defined by the formulae}
\begin{eqnarray}
B_{jk} &=& (1-\delta _{jk})\,\mbox{sin}^{-2}\frac{(j-k)\pi }{n}\,,\\
C_{jk} &=& (1-\delta _{jk})\,\mbox{sin}^{-4}\frac{(j-k)\pi }{n}
\end{eqnarray}
{\em are related to the matrix} $A$ {\em defined above by the equations}
\begin{eqnarray}
B &=& \frac12\left( A^2+2\,A-\sigma _n^{(1)}I\right) , \\
C &=& -\,\frac16\left( B^2-2(2+\sigma _n^{(1)})\,B-\sigma _n^{(2)}I\right)
\end{eqnarray}
{\em where} $I$ {\em is the unit matrix and}
\begin{equation}
\sigma _n^{(1)}=\frac{1}{3}(n^2-1),\quad
\sigma _n^{(2)}=\frac{1}{45}(n^2-1)(n^2+11).
\end{equation}
{\em Thus their eigenvalues corresponding to the eigenvectors} $v^{(s)}$
{\em of components} (3) {\em can be written as follows:}
\begin{eqnarray}
b_s &=& \sigma _n^{(1)}-2s\,(n-s),\qquad s=1,2,\ldots ,n\,,\\
c_s &=& \sigma _n^{(2)}-2s\,(n-s)\,\frac{s(n-s)+2}{3}\,,\qquad
s=1,2,\ldots ,n\,.
\end{eqnarray}

\medskip\noindent
{\bf Theorem 3.} {\em The following sum rules hold}:
\begin{eqnarray}
\sum _{k=1}^{n-1} \mbox{cot}\,\frac{k\pi }n\,\mbox{sin}\, \frac{2
sk\pi }n
&=& n-2s,\qquad s=1,2,\ldots ,n-1\,,\\
\sum _{k=1}^{n-1} \mbox{sin}^{-2}
\frac{k\pi}n\,\mbox{cos}\,\frac{2 sk\pi }n
&=& b_s,\qquad s=1,2,\ldots ,n\,,\\
\sum _{k=1}^{n-1} \mbox{sin}^{-4}\frac{k\pi
}n\,\mbox{cos}\,\frac{2 sk\pi }n
&=& c_s,\qquad s=1,2,\ldots ,n\,, \\
\sum _{k=1}^{n-1} \mbox{sin}^{-2p}\,\frac{k\pi }n &=& \sigma ^{(p)}
\end{eqnarray}
{\em with} $b_s$, $c_s$, $\sigma _n^{(1)}$ {\em
and} $\sigma _n^{(2)}$ {\em defined above and}
\begin{eqnarray}
\sigma _n^{(3)} &=& \sigma _n^{(1)}\,\frac{2n^4+23\,n^2+191}{315}\,,\\
\sigma _n^{(4)} &=& \sigma _n^{(2)}\,\frac{3n^4+10\,n^2+227}{315}\,.
\end{eqnarray}

Theorems 1 and 2 can be easily evinced from the results of [1] or
can be verified by direct computation of the eigenvalue equations
by using of the explicit expression (3) for the eigenvectors. It
should be emphasized that the matrix equations (6) and (7) are
nontrivial consequences of the definitions (4), (5) and (1).
Indeed, the validity of (6) and (7) for the diagonal elements is
easily seen to correspond to the sum rules (14) with $p=1$ and
$p=2$ using the obvious remark that
\begin{equation}
\sum_{k=1,k\neq j}^n\mbox{sin}^{-2p}\frac{(j-k)\pi }n=\sum _{k=1}^{n-1}
\mbox{sin}^{-2p}\,\frac{k\pi }n,\quad j=1,2,\ldots ,n\,. \end{equation}

The validity of (6) for the non-diagonal terms can be verified
explicitly by simple manipulations using the trigonometric identity
\begin{equation}
\mbox{cot}\,\alpha \,\mbox{cot}\,\beta =-\,1-(\mbox{cot}\,\alpha
-\mbox{cot} \,\beta )\,\mbox{cot}\,(\alpha -\beta )\,,
\end{equation}
and the obvious remark
\begin{equation}
\sum _{k=1,k\neq j}^n\mbox{cot}\,\frac{(j-k)\pi }n = \sum
_{k=1}^{n-1} \mbox{cot}\,\frac{k\pi }n=0,\qquad j=1,2,\ldots ,n
\,;
\end{equation}
while the validity of (7) for the non-diagonal terms implies the
nontrivial sum rule
\begin{eqnarray}
&&\sum _{l=1, l\neq j,k}^n\mbox{sin}^{-2}\frac{(j-l)\pi
}n\,\mbox{sin}^{-2}
\frac{(k-l)\pi }n \nonumber \\
&=&2\,(2+\sigma _n^{(1)})\,\mbox{sin}^{-2}\,\frac{(j-k)\pi }n-6\,\mbox{sin}
^{-4}\frac{(j-k)\pi }n,\\
&& j,k =1,2,\ldots ,n,\qquad j\neq k. \nonumber \end{eqnarray}

The first 3 results of Theorem 3, Eqs. (11)--(13), are obtained
from the explicit computation of the eigenvalue equations, while
the last result, Eq. (14), can be verified through an explicit
computation of the traces of $B,\,C,\,BC$ and $C^2$ from (9) and
(10) and from (4), (5) using (17). Computation of the traces of
$AB$ or $AC$ yields instead only special cases of the more general
and quite trivial, formula
\begin{eqnarray}
&& \sum_{k=1,k\neq j}^n\mbox{cot}\,\frac{(j-k)\pi }n\,\left| \mbox{sin}\,
\frac{(j-k)\pi }n\right| ^p=\sum _{k=1}^{n-1} \mbox{cot}\,\frac{k\pi }n\left|
\mbox{sin}\,\frac{k\pi }n\right| ^p=0, \\
&& j=1,2,\ldots ,n, \nonumber \end{eqnarray} that holds for
arbitrary (complex) $p$.

\end{document}